\documentclass[journal=jacsat,manuscript=article]{achemso}
\setkeys{acs}{keywords = true}
\usepackage[version=3]{mhchem} 
\pdfoutput=1
\usepackage{rotating}
\usepackage{multirow}
\usepackage{amsthm}
\usepackage{amsmath}
\usepackage{diagbox}
\usepackage{bbm, dsfont}
\usepackage{graphicx}
\usepackage{booktabs}
\usepackage{siunitx}
\usepackage{hyperref}
\usepackage{caption}
\DeclareSIUnit{\bps}{bps}
\DeclareSIUnit{\cps}{cps}

\SectionNumbersOn
\usepackage{rotating}
\usepackage{multirow}
\usepackage{amsthm}
\usepackage{amsmath}
\usepackage{diagbox}
\usepackage{bbm, dsfont}
\usepackage{graphicx}
\usepackage{booktabs}
\usepackage{siunitx}
\usepackage[normalem]{ulem}
\usepackage{caption}
\DeclareSIUnit{\bps}{bps}
\DeclareSIUnit{\cps}{cps}
\author{Giovanni V. Resta}
\email{giovanni.resta@unige.ch}
\affiliation[Univeristy of Geneva]
{Group of Applied Physics, Univeristy of Geneva, Rue de l'Ecole-de-Médecine 20, CH-1211, Genève, Switzerland}
\alsoaffiliation[ID Quantique SA]
{ID Quantique SA, Rue Eugène-Marziano 25, CH-1227, Genève, Switzerland}
\altaffiliation{These authors contributed equally}
\author{Lorenzo Stasi}
\affiliation[ID Quantique SA]
{ID Quantique SA, Rue Eugène-Marziano 25, CH-1227, Genève, Switzerland}
\alsoaffiliation[Univeristy of Geneva]
{Group of Applied Physics, Univeristy of Geneva, Rue de l'Ecole-de-Médecine 20, CH-1211, Genève, Switzerland}
\altaffiliation{These authors contributed equally}
\author{Matthieu Perrenoud}
\altaffiliation{These authors contributed equally}
\affiliation[Univeristy of Geneva]
{Group of Applied Physics, Univeristy of Geneva, Rue de l'Ecole-de-Médecine 20, CH-1211, Genève, Switzerland}
\author{Sylvain El-Khoury}
\affiliation[ID Quantique SA]
{ID Quantique SA, Rue Eugène-Marziano 25, CH-1227, Genève, Switzerland}
\author{Tiff Brydges}
\affiliation[Univeristy of Geneva]
{Group of Applied Physics, Univeristy of Geneva, Rue de l'Ecole-de-Médecine 20, CH-1211, Genève, Switzerland}
\author{Rob Thew}
\affiliation[Univeristy of Geneva]
{Group of Applied Physics, Univeristy of Geneva, Rue de l'Ecole-de-Médecine 20, CH-1211, Genève, Switzerland}
\author{Hugo Zbinden}
\affiliation[Univeristy of Geneva]
{Group of Applied Physics, Univeristy of Geneva, Rue de l'Ecole-de-Médecine 20, CH-1211, Genève, Switzerland}
\author{Félix Bussières}
\affiliation[ID Quantique SA]
{ID Quantique SA, Rue Eugène-Marziano 25, CH-1227, Genève, Switzerland}

\title[GHz detection rates and dynamic photon-number resolution with superconducting nanowire arrays]{GHz detection rates and dynamic photon-number resolution with superconducting nanowire arrays}


\begin{document}

\begin{tocentry}

\includegraphics[width = 82.5mm]{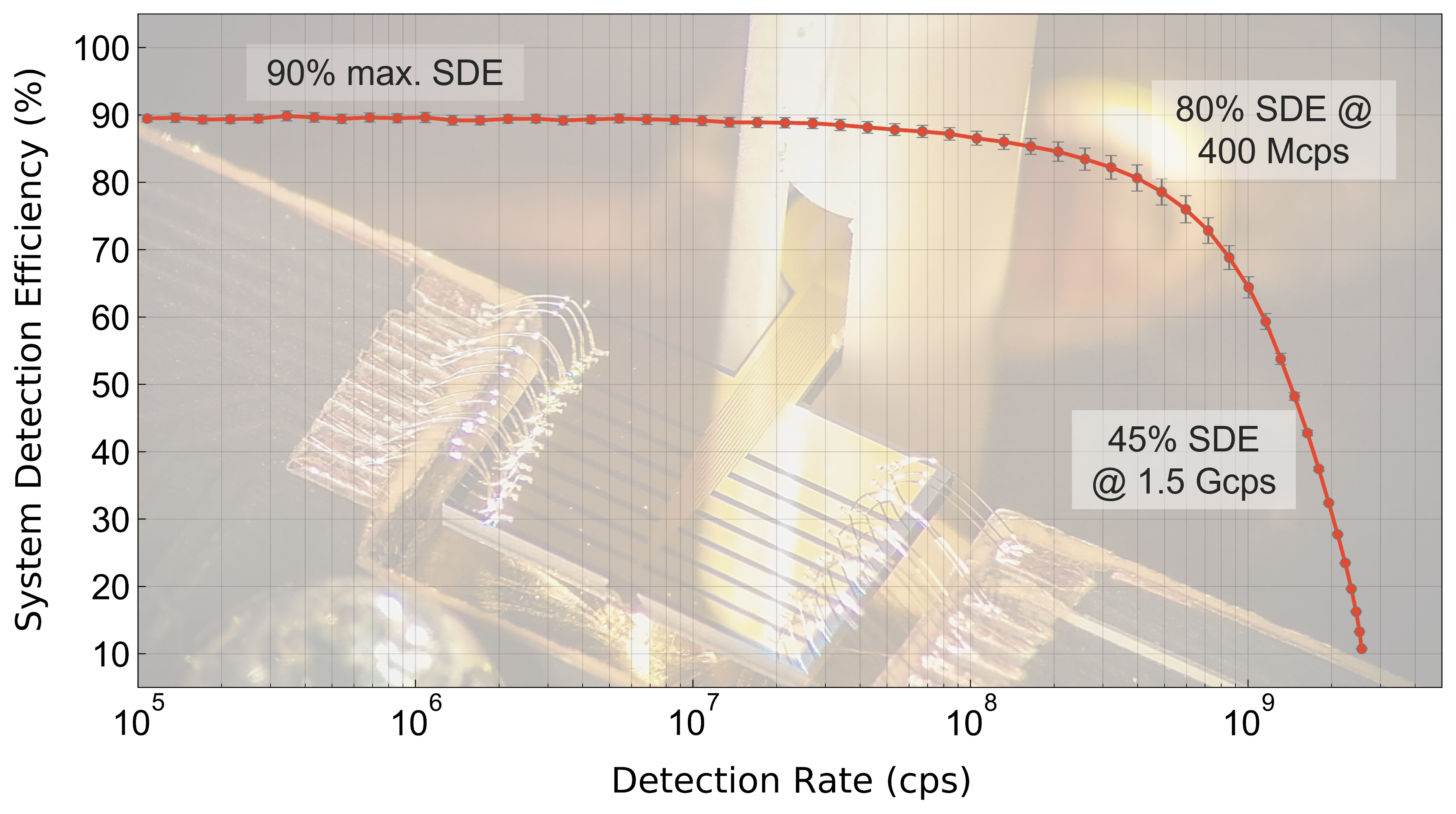}

\end{tocentry}

\begin{abstract}
Superconducting-nanowire single-photon detectors (SNSPDs) have enabled the realization of several quantum optics technologies thanks to their high system detection efficiency (SDE), low dark-counts, and fast recovery time. However, the widespread use of linear optical quantum computing, quasi-deterministic single photon sources and quantum repeaters requires even faster detectors that can also distinguish between different photon number states. Here, we present an SNSPD array composed of 14 independent pixels, achieving a SDE of 90\% in the telecom band. By reading each pixel of the array independently, we show detection of telecom photons at \SI{1.5}{\giga \hertz} with 45\% absolute SDE. We exploit the dynamic photon number resolution (PNR) of the array to demonstrate accurate state reconstruction for a wide range of light inputs, including operation with long-duration light pulses, as obtained with some cavity-based sources. We show 2\nobreakdash-photon and 3\nobreakdash-photon fidelities of $74\%$ and $57\%$ respectively, which represent state-of-the-art results for fiber-coupled SNSPDs.
\end{abstract}

\section{Introduction}\label{intro}
Optical quantum technologies crucially rely on the high-speed detection of single photons with high efficiency, low noise and high timing resolution at a variety of wavelengths, ranging from the visible to the near-IR. Moreover, the ability to resolve photon-number states in an optical pulse is a critical requirement for linear optic quantum computing (LOQC)~\cite{knill2001scheme, slussarenko2019photonic}, gaussian boson sampling (GBS)~\cite{wang2019boson, arrazola2021quantum, madsen2022quantum}, quantum network protocols based on quantum repeaters~\cite{sangouard2011quantum}, quantum metrology~\cite{von2019quantum}, and to achieve quasi\nobreakdash-deterministic heralded single\nobreakdash-photon sources~\cite{meyer2020single, sempere2022reducing, davis2022improved, stasi2022enhanced}.

Since their inception, single-pixel superconducting nanowire single-photon detectors (SNSPDs)\cite{gol2001picosecond} have had a profound impact in the field of optical quantum information processing and have demonstrated remarkable performances, including near unity system detection efficiency \cite{reddy2020superconducting, chang2021detecting}, timing jitter below \SI{10}{\pico\second} \cite{esmaeil2020efficient, korzh2020demonstration}, low dark counts \cite{shibata2015ultimate,boaron2018secure}, and recovery times on the order of tens of nanoseconds, typically leading to maximum count rates of tens of MHz. However, a wide range of applications, from quantum-key distribution (QKD)~\cite{boaron2018secure}, to integrated single-photon sources such as those using quantum dots~\cite{wein2022photon}, spontaneous four-wave mixing (SFWM)~\cite{samara2019high,ma2020ultrabright, steiner2021ultrabright} or spontaneous parametric down-conversion (SPDC)~\cite{cabrejo2022ghz}, would greatly benefit from detectors that could provide a considerably higher detection rate while maintaining a high efficiency. Moreover, in their simplest implementation, an SNSPD is a meandered nanowire which acts as a \textit{`bucket detector'}, \textit{i.e.} it is only able to distinguish \textit{`zero'} or \textit{`many'} photons, and lacks photon-number resolution (PNR) capabilities. 

In order to achieve PNR capability with SNSPDs, spatial-multiplexing designs consisting of several pixels, connected in parallel or in series, have been investigated~\cite{divochiy2008superconducting, marsili2009superconducting, jahanmirinejad2012photon, mattioli2016photon, stasi2022high}, and operation of parallel designs at detection rates up to \SI{200}{\mega \cps} have also been demonstrated using an additional anti-latching architecture \cite{perrenoud2021operation, stasi2022high}. Time-multiplexing has also been utilized in waveguide-integrated SNSPDs arrays, with discrimination of high photon number states already being demonstrated \cite{cheng2023100}, at the cost of limiting the detection rate of the entire array to a maximum of $\sim$\SI{10}{\mega \cps}. 

However, both series and parallel SNSPDs are still limited in the maximum detection rates achievable and can only resolve photon numbers with light pulse duration of up to a few hundred picoseconds at most~\cite{divochiy2008superconducting}, a limitation that could severely limit the use of such detectors. In fact, photons obtained from cavity-enhanced SPDC~\cite{Pomarico_2012,Monteiro:14,slattery2019background} or SFWM~\cite{steiner2021ultrabright, brydges2022integrated}, often have long coherence times on the order of a few nanoseconds. Such photons can be used in LOQC \cite{slussarenko2019photonic}, GBS \cite{arrazola2021quantum, madsen2022quantum}, and in quantum repeater protocols for long distance communication~\cite{sangouard2011quantum, clausen2014source}, which generally require very narrow bandwidth photons in order to match those of quantum memories. Independent multi-pixel SNSPD arrays have been investigated to remove the limitations of series and parallel designs in terms of maximum detection rate, at the cost of a more complex read-out\cite{dauler2007multi, zhang201916, grunenfelder2023fast, craiciu2023high}. However, a full analysis of the PNR capabilities of independent multi-pixel arrays and a demonstration of operation with long light pulses is crucially still lacking. 

Here, we present an SNSPD array composed of 14 independent pixels, that is able to be operated in free-running mode and simultaneously deliver high performance in terms of system detection efficiency, jitter, recovery-time and maximum count rate. The detector also supports dynamic PNR capability, where \textit{dynamic} means that there is no limitation caused by the duration of the light pulse being analyzed.

\section{Results and Discussion}\label{sect:results}
\subsection{Detector architecture and fabrication}\label{subsect:detector}

\begin{figure}[ht!]
	\includegraphics[width = 1.
	\columnwidth]{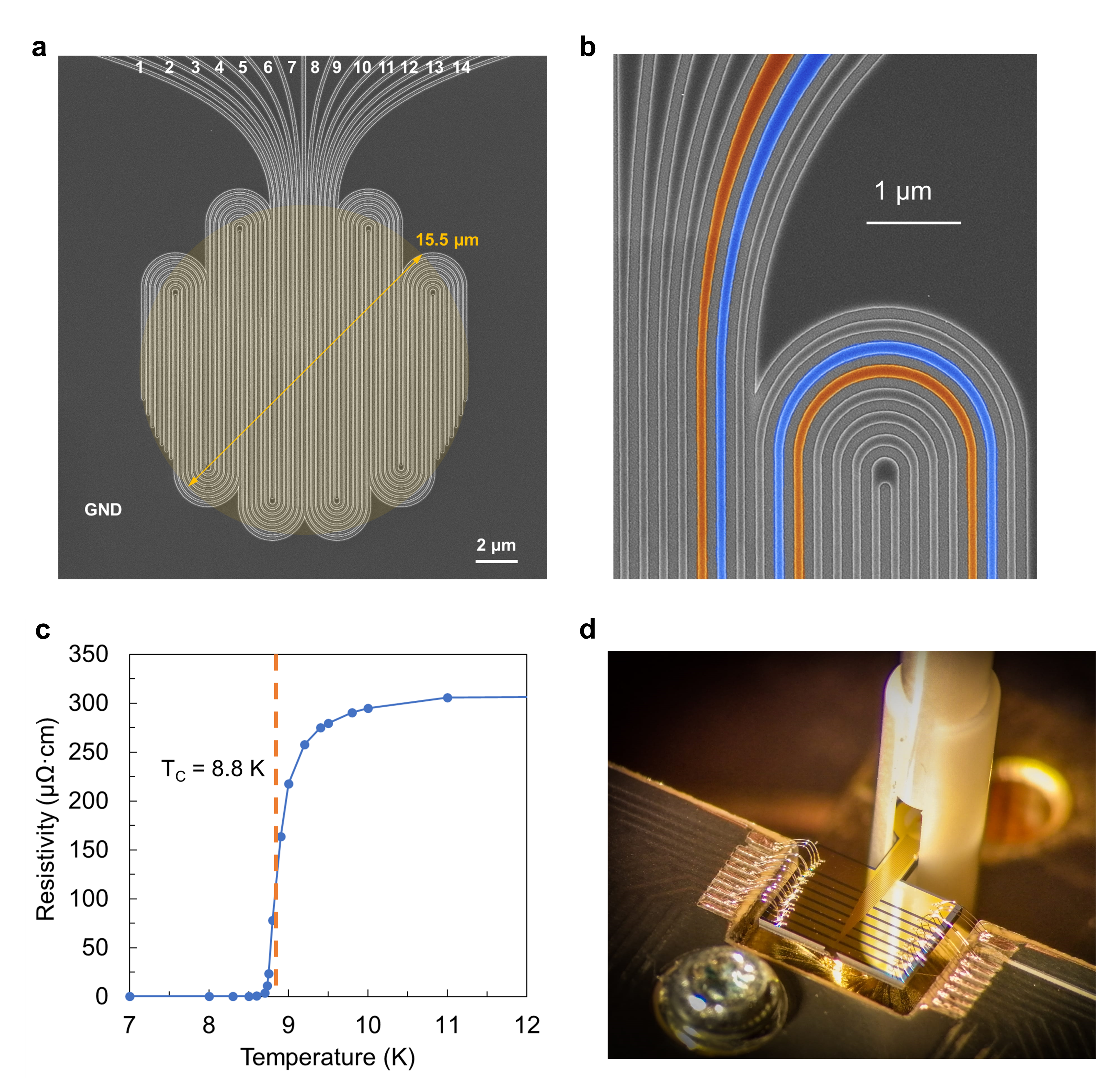}
	\caption{ \textbf{14-pixel SNSPD array.} \textbf{a,} SEM image of the 14-pixel detector array. The detector covers a circular area with a diameter of $\sim$\SI{15.5}{\um} which, when coupled to an SMF-28 single mode optical fiber, is enough to collect more than $99\%$ of the incoming light. \textbf{b,} Recolored SEM image showing the design of the turns and highlighting the path of two adjacent pixels. The width of the nanowires is \SI{100}{\nm} with a fill factor of 50\%. \textbf{c,} Plot of resistivity vs temperature, showing the critical temperature of the NbTiN superconducting film used to fabricate the SNSPD array. \textbf{d,} Image of the packaged detector, showing the coupling with the optical fiber and the wire-bonding to the PCB at \SI{0.8}{K}.}
 \label{fig:device}
\end{figure}

The multi-pixel SNSPD array is designed such that high efficiency, low jitter, high maximum count rate and photon-number resolving capabilities are simultaneously achieved. The multi-pixel SNSPD array covers an area of roughly \SI{200}{\um^2}, and is composed of 14 independent pixels arranged in an interleaved geometry in order to guarantee a uniform light distribution, as shown in Fig. \ref{fig:device}(a). The ground-plane (indicated by GND in Fig. \ref{fig:device}(a)), surrounds the detector and is connected to the same electrical ground as the entire cryostat.

The nanowires have a width of \SI{100}{\nm} with a spacing of \SI{100}{\nm}, thus giving a total fill-factor (FF) in the illuminated area of 0.5, as shown in Fig. \ref{fig:device}(b). In order to minimize the length of the turns, and thus of each nanowire, we opted for a design where 7 nanowires cover the right side of the array and 7 nanowires cover the left side of the array. The effects of this separation (which we refer to as a semi-interleaved design) will become evident when testing each pixel individually (see Fig. \ref{fig:1pix}).

The device is integrated into an optical cavity, designed to maximize photon absorption at \SI{1550}{\nm}. We use niobium-titanium nitride (NbTiN), sputtered from a NbTi target in a nitrogen-rich atmosphere, as the superconducting material (see Supporting Information for a detailed description of the fabrication process). The superconducting film has a thickness of around \SI{9}{\nm} and exhibits a critical temperature ($T_{c}$) of \SI{8.8}{\kelvin} (see Fig. \ref{fig:device}(c)). The device is wire-bonded to a custom-made PCB and placed in a closed cycle cryostat (ID281, ID Quantique), at \SI{0.8}{\K}. The array is fiber-coupled to an SMF-28 single-mode optical fiber with a self-aligned packaging scheme similar to the one described in Ref. \cite{miller2011compact} (as shown in Fig. \ref{fig:device}(d)). 

\subsection{Single-pixel characterization}\label{subsect:1pix}

\begin{figure}[ht!]
	\includegraphics[width = 1.
	\columnwidth]{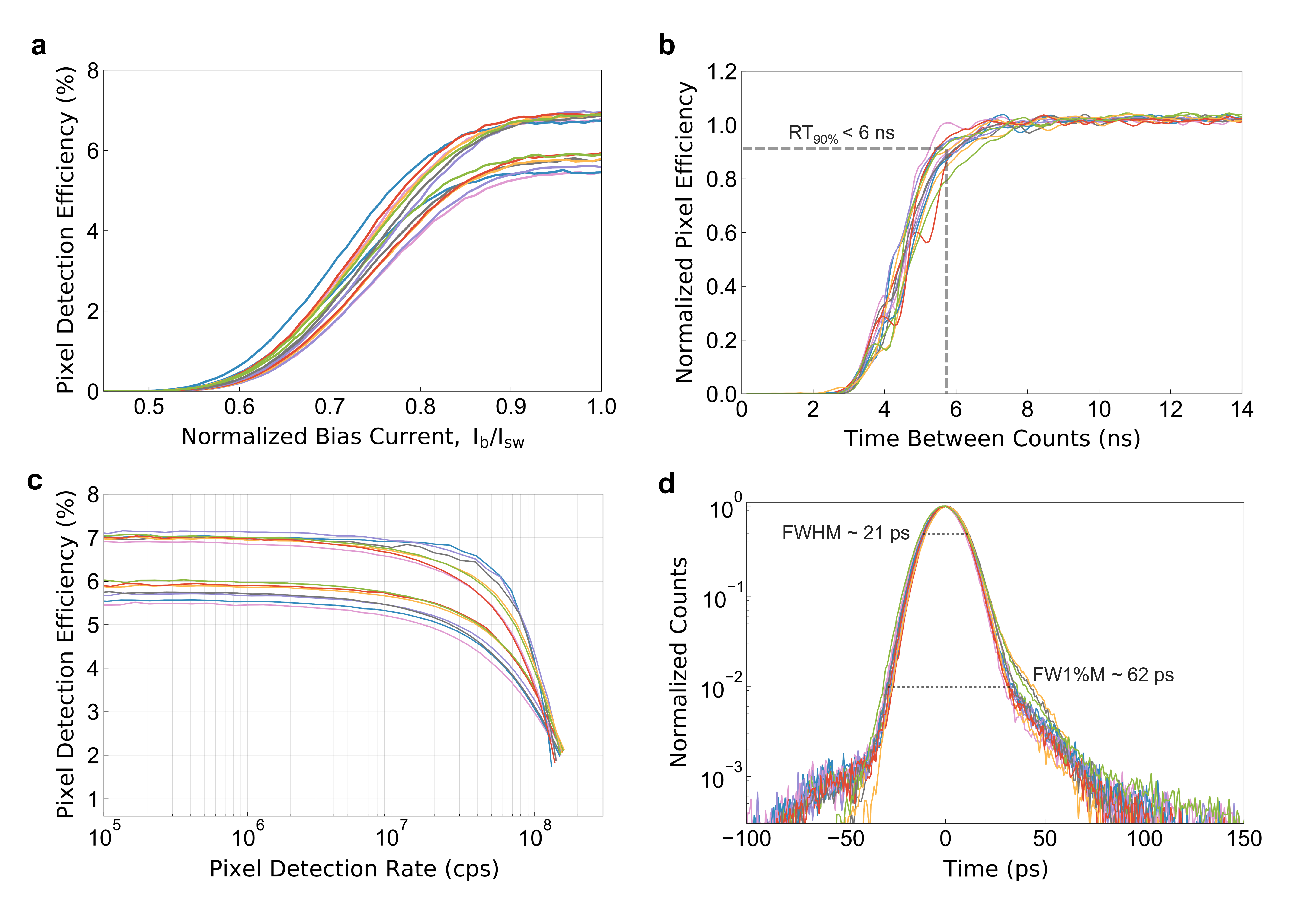}
	\caption{\label{fig:1pix} \textbf{Single-pixel characterizations.} \textbf{a,} Single-pixel detection efficiencies measured at \SI{1550}{\nm} as a function of the bias current. \textbf{b,} Recovery time measurement for the 14 individual pixels. The  RT$_{90\%}$ is lower than \SI{6}{ns} for all pixels. \textbf{c,} Count rate curve for each pixel measured with a CW laser at \SI{1550}{\nm} and at a fixed bias current. Each pixel has a maximum count rate of $\sim$\SI{100}{\mega\cps}. \textbf{d,} Jitter measurement for the 14 pixels, showing an average FWHM jitter of \SI{21}{\ps} and great uniformity amongst all pixels. All measurements are performed individually on each pixel of the array.}
\end{figure}

We verify the performances of each individual pixel in terms of efficiency, recovery time, count rate, and jitter (a detailed description of the read-out scheme and experimental set-ups used for these measurements can be found in the Supporting Information). We acquire the single pixel efficiencies by illuminating the array with a continuous-wave (CW) laser at \SI{1550}{\nm} and sweeping the bias current ($I_{b}$) of each pixel, while all other pixels are not biased. The currents are normalized with respect to the switching current ($I_{sw}$) of each pixel, which we define as the current at which the dark count rate (DCR) exceeds \SI{10}{\cps}. In the case of multipixel arrays, where the number of input photons on each pixel cannot be controlled, the single pixel detection efficiencies are a representation of the spatial distribution of the input light on the array. As can be seen in  Fig. \ref{fig:1pix}(a), all 14 pixels exhibit a plateau, \textit{i.e.} region of saturated internal efficiency, making it possible to tune $I_{b}$ without affecting the performances of the detector. By summing up the maximum single-pixel efficiencies, we estimate a total efficiency for the entire array of $\sim 89\%$. This estimation was verified experimentally, as presented later in Fig. \ref{fig:CR}. Due to our semi-interleaved design, a minor misalignment between the fiber and the center of the array resulted in a slightly uneven illumination of the detector, causing one set of 7 pixels to present a higher efficiency than the other set (see Fig. \ref{fig:1pix}(a)).

The detector covers the same area as a conventional single-pixel SNSPD (around \SI{200}{\um^2}), thus the length of each nanowire is greatly reduced, and so is their kinetic inductance ($L_k$), allowing for much faster operation. Fig. \ref{fig:1pix}(b) shows the normalized pixel detection efficiency as a function of the inter-arrival time between two consecutive photon detections. From this curve we can extract the recovery time of each pixel, which we define as the time needed, after a detection, to recover $90\%$ of the maximum pixel efficiency ($RT_{90\%}$). For all 14 pixels the $RT_{90\%}$ is lower than \SI{6}{\ns}, which is at least 6$\times$ faster than a typical single-pixel SNSPDs \cite{autebert2020direct}, and directly translates into the capability to reach high count rates, as can be seen in Fig. \ref{fig:1pix}(c). Here, we illuminate the array with a CW laser at \SI{1550}{\nm} and measure the pixel detection efficiency as a function of the pixel detection rate, with each pixel biased close to $I_{sw}$. A commonly used metric for evaluating the maximum count rate (MCR) of an SNSPD is to extract the detection rate at 50\% value of the maximum efficiency\cite{craiciu2023high}. As seen in Fig. \ref{fig:1pix}(c) each pixel has a MCR of $\sim$\SI{100}{\mega\hertz}, thus showing the potential of achieving GHz counting rates when simultaneously operating the entire array. The results presented in Fig. \ref{fig:CR} experimentally demonstrate this capability.

We characterize the jitter of each pixel at \SI{1550}{\nano \meter} and at low count rates (\textit{i.e.} $\sim$\SI{100}{\kilo \cps} for each pixel), using a commercially available time-correlated single photon counting (TCSPC) module with a constant-fraction discriminator. We obtain an average full width at half maximum (FWHM) jitter of \SI{21}{\pico\second} (without any de-convolution) and an average full width at 1\% maximum (FW1\%M) jitter of \SI{62}{\pico\second} for currents close to $I_{sw}$ (see Fig. \ref{fig:1pix}(d)). These average jitter values are also representing the timing performances of the entire array.

\subsection{Thermal crosstalk analysis}\label{subsect:crosstalk}

\begin{figure}[ht!]
	\includegraphics[width = 1.
	\columnwidth]{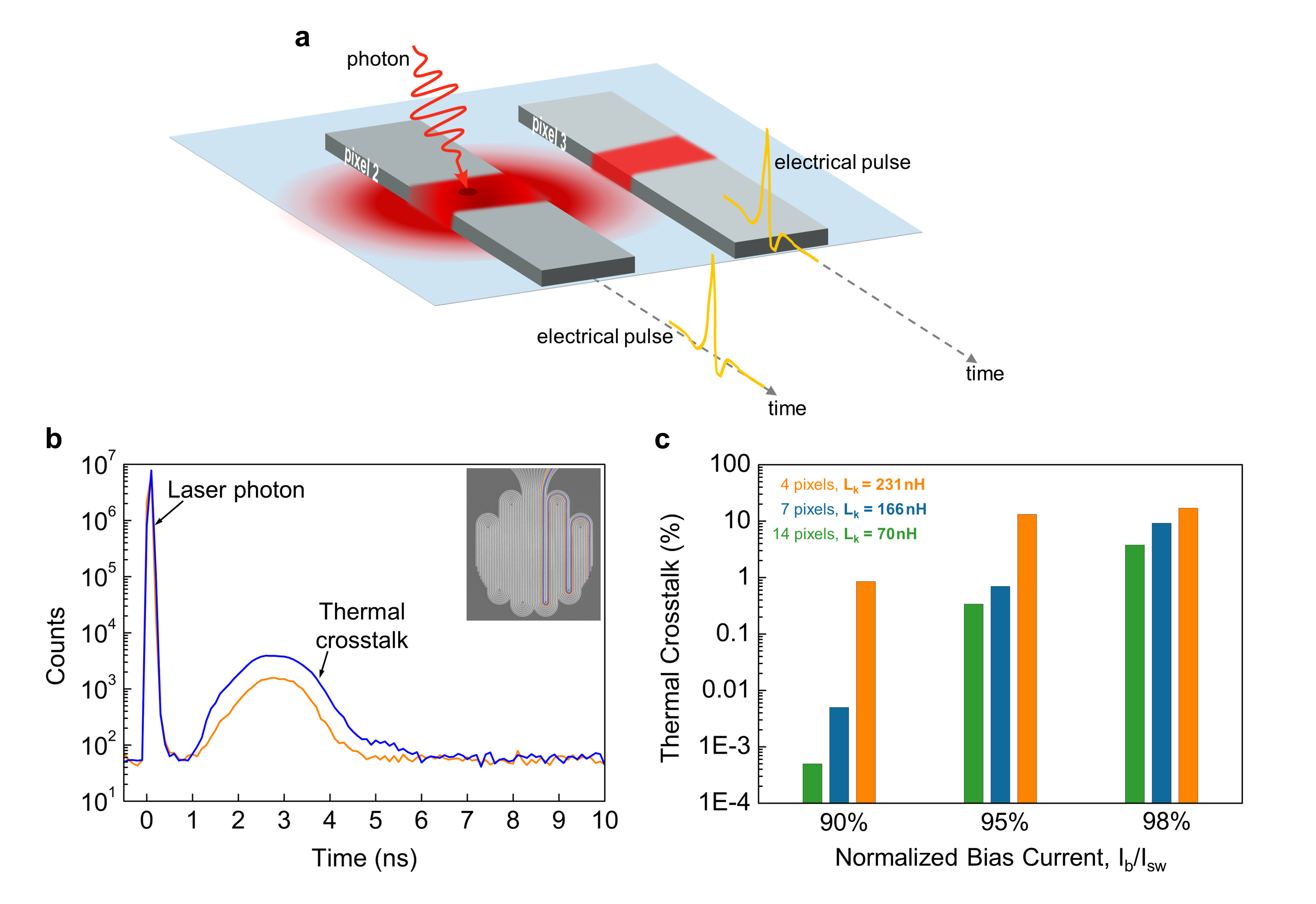}
	\caption{\label{fig:crosstalk} \textbf{Thermal crosstalk analysis.} \textbf{a,} Schematic depiction of the mechanism from which thermal crosstalk originates in the multi-pixel interleaved geometry. Note how two pulses separated in time are produced upon the absorption of a single photon. \textbf{b,} Count histogram of two adjacent pixels illuminated with a picosecond laser. Thermal crosstalk appears roughly \SI{1}{\ns} after a photon detection in the adjacent nanowire and lasts for $\sim$\SI{4}{\ns}. It can clearly be seen that the two pixels have a different thermal crosstalk probability, as the integral of the thermal crosstalk counts is different, since they are operated at a different ratio of $I_b/I_{sw}$. The SEM inset highlights the adjacent nanowires measured in the experiment. \textbf{c,} Measured thermal crosstalk probability between adjacent pixels as a function of bias current for 3 designs covering the same area with 4, 7 and 14 pixels, corresponding to a kinetic inductance of 231~nH, 166~nH and 70nH respectively. Both bias current and kinetic inductance have a strong impact on the probability of generating thermal crosstalk.}
\end{figure}

Before operating the array in our experiments, and thus biasing all the pixels of the array simultaneously, we analyze the effect of thermal crosstalk between the pixels. A schematic depiction of the mechanism responsible for thermal crosstalk is presented in Fig. \ref{fig:crosstalk}(a). The absorption of a single photon in one of the pixels of the SNSPD will cause the formation of a resistive hot-spot (broken cooper pairs), and lead to a local suppression of superconductivity in the nanowire. The resistance created, which we refer to as hotspot resistance($R_{hs}$), and is on the order of a few k$\Omega$, will divert the bias current in the read-out electronics and generate an output pulse, which is then used to count the absorbed photons. During this process the nanowire locally heats up, due to Joule heating, and this heat diffuses into the substrate  \cite{yang2007modeling}. It is then possible, that this local increase in temperature triggers a false detection in an adjacent nanowire, thus generating a second electrical pulse, despite the absence of an absorbed photon on that pixel. This second pulse will be delayed in time with respect to the first one, as seen in the measurements displayed in Fig. \ref{fig:crosstalk}(b). Here a picosecond pulsed laser is used to illuminate the SNSPD array and the counts between two adjacent pixels (as shown in the inset) are analyzed. We can see the effect of thermal crosstalk already occurring after a delay of \SI{1}{\ns} and for a duration of roughly \SI{4}{\ns} (see Fig. \ref{fig:crosstalk}(b)). From such experimentally measured histograms, the thermal crosstalk probability is directly computed by dividing the sum of the counts generated by thermal crosstalk in one pixel by the counts generated on the neighbouring pixel by the laser. 

The thermal crosstalk probability in adjacent nanowires is a function of their spatial separation, bias current, and kinetic inductance ($L_k$). The value of $L_k$ determines the amount of heat that is generated by the nanowire since, together with $R_{hs}$, it imposes a time constant ($\tau = L_k/R_{hs}$) on the process for the current leaving the pixel and going to the readout electronics. As a consequence, the thermal crosstalk probability is also impacted by the current flowing in each pixel, which can result in a non-uniform crosstalk distribution among the pixels (as shown in Fig.~\ref{fig:crosstalk}(b)). 

We study the dependence of the thermal crosstalk with respect to $L_k$ and bias current of the pixels in different interleaved geometries, with the width of the nanowires and their separation fixed to \SI{100}{\nm} (see Fig.~\ref{fig:crosstalk}(c)). The $L_k$ is varied by fabricating detectors with the same total area, but different numbers of pixels, hence the more pixels, the shorter their individual length. Measurements are taken with the pixels biased at three different $I_b/I_{sw}$ ratios. For the 14-pixel SNSPD design presented in this work, we obtain thermal crosstalk probabilities between adjacent pixels below 0.5\% when biasing the pixels at $0.95\cdot I_{sw}$ and, when reducing the bias current to $0.9\cdot I_{sw}$, the thermal crosstalk probability between adjacent pixels drops to below $10^{-3}$\%. Multiplying this value by the total number of combinations of adjacent nanowires, gives a crosstalk probability for the entire array of $\sim0.015\%$. Thus, we conclude that a bias current of $0.9\cdot I_{sw}$ for each pixel represents an optimal biasing point for the array, since we can preserve high efficiency, \textit{i.e.}~operate the pixels on the plateau region, while minimizing the thermal crosstalk probability.

\subsection{Array total efficiency and detection rate}\label{subsect:array}

\begin{figure}[ht!]
	\includegraphics[width = 0.9
	\columnwidth]{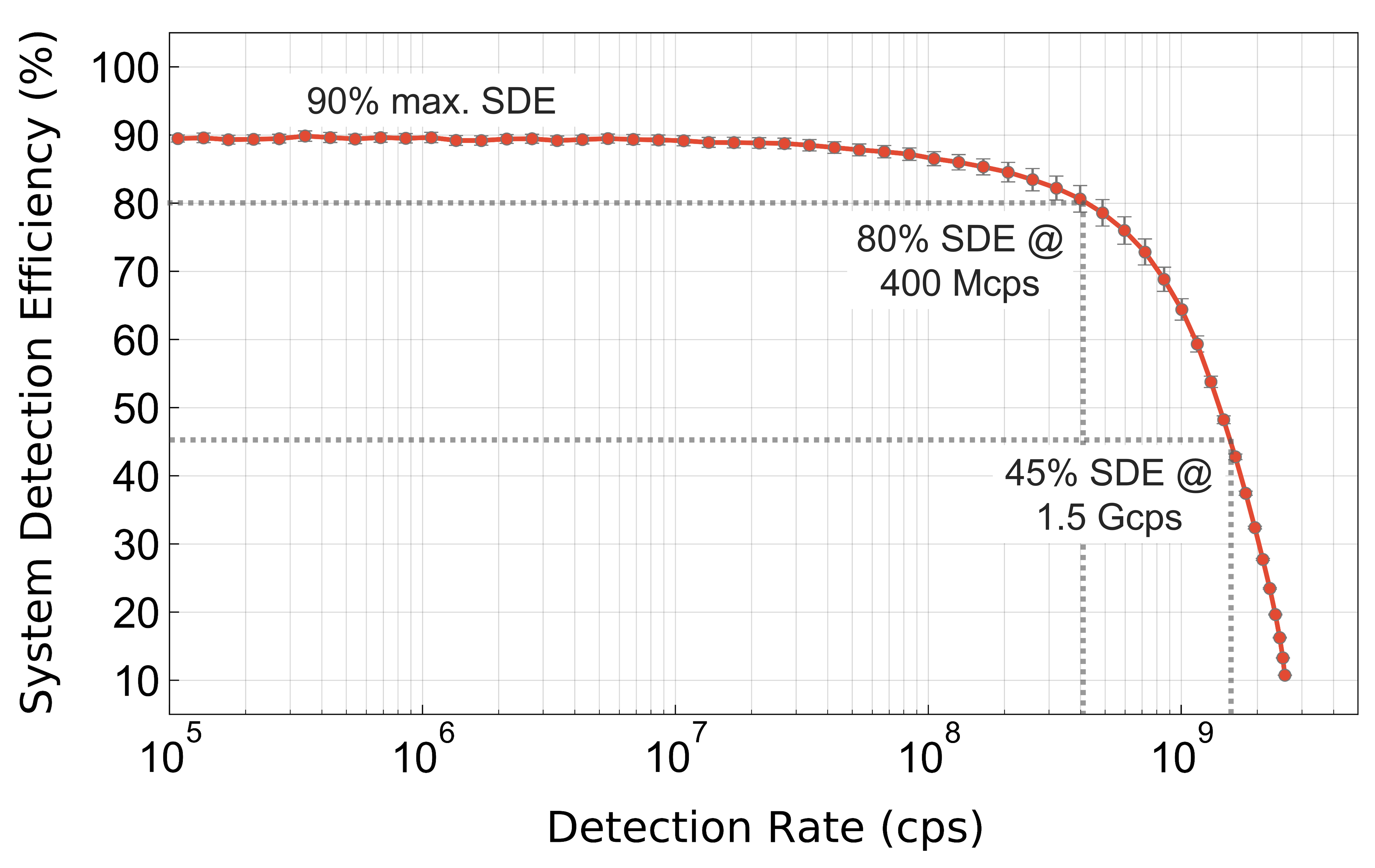}
	\caption{\label{fig:CR} \textbf{SDE as a function of detection rate.} Measured SDE vs detection rate curve acquired with all 14 pixels active. The maximum efficiency of the array is $89.5\%\pm0.7\%$ and the MCR is \SI{1.5}{\giga\cps}.}
\end{figure}

After characterizing each pixel individually, and verifying the negligible impact of thermal crosstalk, we characterize the performance of the array in terms of total system detection efficiency (SDE) and detection rate capability in free running mode, as shown in Fig. \ref{fig:CR}. For this measurement we illuminate the array with a CW laser at \SI{1550}{\nm} and measure the average efficiency per photon, \textit{i.e.} the inter-arrival time between consecutive photons is not fixed. All 14 pixels are biased at $I_b = 0.9\cdot I_{sw}$, which allows the thermal crosstalk to be minimized while maintaining high pixel detection efficiency, and read-out simultaneously using 3 time controllers (ID1000, ID Quantique). The sum of the counts on each pixel gives the total SDE of the array, and the DCR, measured with the source plugged in the system, is below \SI{150}{\cps}. As can be seen in Fig. \ref{fig:CR}, we measure a maximum SDE of $89.5\%\pm0.7\%$ for detection rates up to \SI{10}{\mega\cps}, and the SDE remains above 80\% up to \SI{400}{\mega\cps}. This is consistent with the total efficiency estimation of $89\%$ obtained from the single-pixel characterization. As the count rates increases, the probability that photons are absorbed during the recovery time of each nanowire also increases, and thus the average SDE per photon begins to reduce. The MCR of the entire array, with all pixels operated simultaneously, is \SI{1.5}{\giga\cps} (as can be seen in Fig. \ref{fig:CR}. Thanks to the latch-free operation, we can also operate the detector at even faster detection rates, and we measure an average SDE per photon of 27\% at \SI{2.1}{\giga\cps}. 

\subsection{Photon-number resolving capabilities} \label{subsect:pnr}

We finally demonstrate the potential of our SNSPD array for use in quantum optics experiments, operated as a photon\nobreakdash-number resolving (PNR) detector, by measuring and reconstructing the photon statistics of poissonian light. The measured photon\nobreakdash-counting statistics, ${Q_n}$, namely the probability of measuring an $n$-click event by a PNR detector, can be directly linked to the incoming light statistics, $S_m$ (the probability to have $m$ photons in the fibre connected to the whole SNSPD system), via ${Q_n} = \sum_{m=0}^{\infty} P_{nm} \cdot S_m$, where ${P_{nm}}$ is the probability of registering an $n$-click when $m$ photons are incident. We refer to the ensemble of ${P_{nm}}$ values as the probability matrix $\mathbf{P}$, and the elements are estimated following Ref.~\cite{fitch2003photon}, with the \textbf{P} matrix (truncated to 8-photon events) presented in table \ref{table:povm} (see Supporting Information for more details). From the diagonal elements of the \textbf{P} matrix, \textit{i.e.} the $P_{nn}$ elements, we can retrieve the probabilities to correctly detect an incoming photon-number state, which we refer to as photon fidelities.

\begin{figure}[ht!]
	\includegraphics[width = 1.
	\columnwidth]{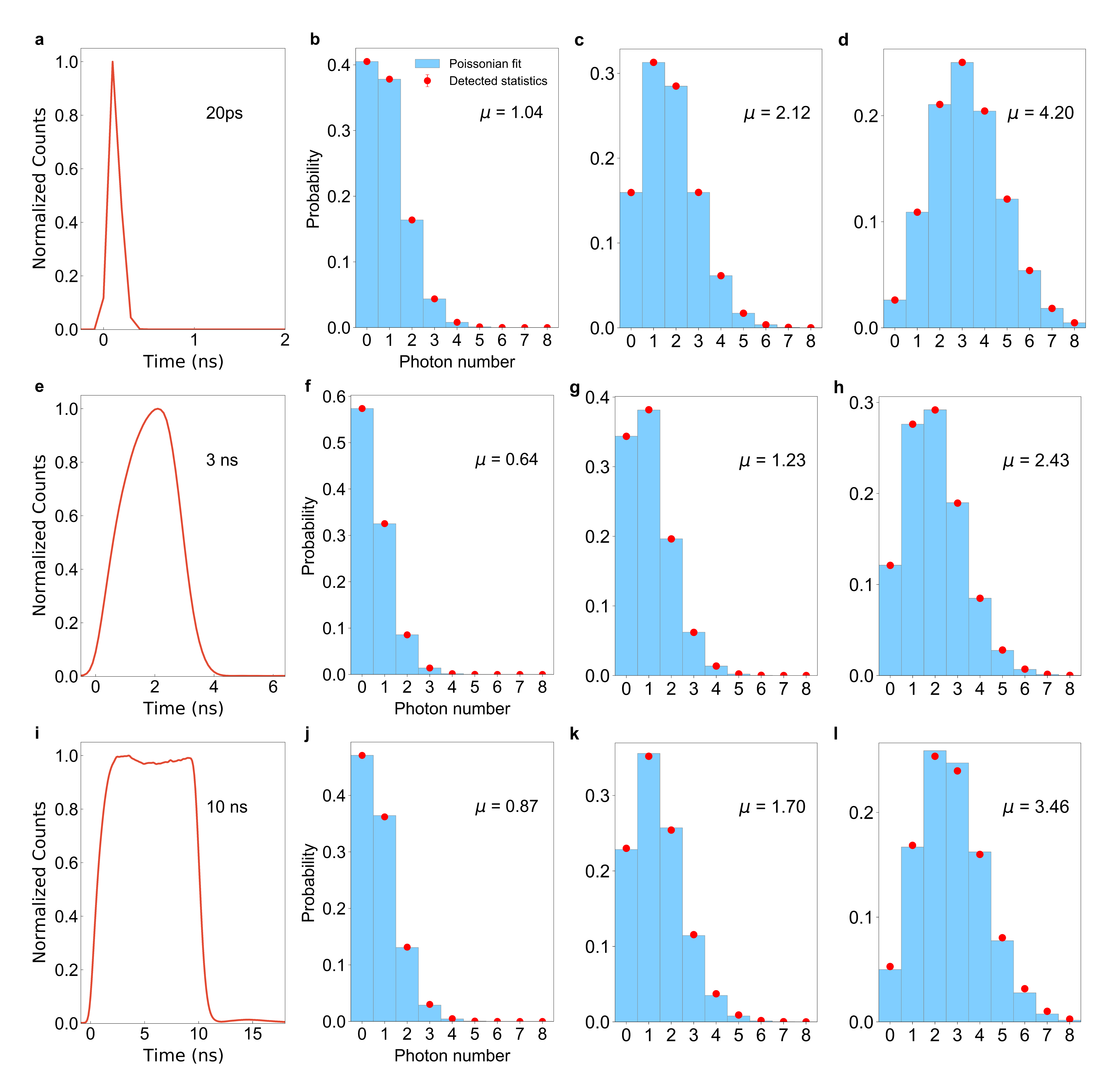}
	\caption{\label{fig:PNR} \textbf{Photon-number resolution and state reconstruction with long pulses of light.} \textbf{a,e,i} Estimated profile retrieved from the counting histogram of a single pixel for \SI{20}{\pico\second} (a), \SI{3}{\nano\second} (e), and \SI{10}{\nano\second} (i) light-pulses. \textbf{b-d, f-h, j-l} Experimental and retrieved photon\nobreakdash-counting statistics under Poissonian light illumination of \SI{20}{\pico\second} (b-d), \SI{3}{\nano\second} (f-h) and \SI{10}{\nano\second} (j-l) light pulses. The retrieved $\mu$ are also shown in the figures. Note that for the \SI{3}{\nano\second} and \SI{10}{\nano\second} light pulses the  $\mu$ could not be precisely set, as in the case of the pulsed-laser, due to the finite extinction-ratio of the amplitude modulator.}
\end{figure}

To confirm the correctness of the detector $\mathbf{P}$ matrix, poissonian light at \SI{1}{\mega\hertz} with \SI{20}{\pico\second}-long pulses and a mean\nobreakdash-photon\nobreakdash-number per pulse ($\mu$) of 1, 2, and 4 is sent to the detector. The generated electrical signals from the pixels of the array are acquired and post-processed to obtain the experimental photon-counting statistics ${Q_n}$ (see the Supporting information  for further details), which are then used to retrieve the $\mu$ of the poissonian light. The results are shown in Fig.~\ref{fig:PNR}(a-d). The retrieved $\mu$ lie within a $5\%$ margin compared to the expected value, which can be attributed to the systematic uncertainty of the power-meter, thus supporting the correctness of the $\mathbf{P}$ matrix. The detector has 2\nobreakdash-photon and 3\nobreakdash-photon fidelities of $74\%$ and $57\%$, respectively, which represents a significant step forward with respect to the best reported values in literature for fiber-coupled SNSPDs, \textit{e.g.}~$48\%$ for 2\nobreakdash-photon fidelity and $9\%$ for a 3\nobreakdash-photon state\cite{stasi2022high}. To illustrate the impact this can have in a concrete example, coupling this 14-pixel detector with an heralded single-photon source to suppress multi-photon pair emission would reduce the zero-time second-order autocorrelation ($g^{(2)}(0)$) of the heralded single-photon by more than 67\% when compared to a heralding event that would use a 90\% efficient bucket-detector instead (assuming a heralding transmission of 0.95).

As previously mentioned, many promising communication protocols based on quantum repeaters require the use of heralded single photon sources along with quantum memories for storage of entanglement \cite{Sangouard:2011}. These systems require narrow bandwidth photons, corresponding to photons with long coherence times up to several nanoseconds \cite{Businger:2020,Askarani:2021}, which we refer to as the long light-pulse regime. Such a regime is also relevant for implementations of LOQC or GBS using squeezed states of light generated from cavity-enhanced SPDC~\cite{madsen2022quantum} or SFWM sources~\cite{ma2020ultrabright}. As such, photon-number resolving detectors which can extract photon-number information from long light pulses is likely to become crucial in the implementation of these systems. However, parallel and series SNSPDs and impedance matching taper SNSPDs~\cite{zhu2020resolving} can only operate with short light pulses (hundreds of picoseconds), due to their intrinsic detection mechanism. We test the performance of our detector in the long light-pulse regime, by sending coherent light pulses of \SI{3}{\nano\second} and \SI{10}{\nano\second} produced by amplitude modulating a \SI{1550}{nm} CW laser, see Fig.~\ref{fig:PNR}(e-h) and Fig.~\ref{fig:PNR}(i-l). Even in this regime, we are able to correctly reconstruct the input light statistics for $\mu$ up to 3.4, thus proving the dynamic PNR capability of our device, and enabling the use of multipixel SNSPD arrays in the long light-pulse regime. 

\begin{table}
\footnotesize
\begin{center}
\caption{Truncated $\mathbf{P}$ matrix (up to 8-photon events) of the 14-pixel SNSPD array.}\label{table:povm}%
\begin{tabular}{c c | c c c c c c c c c}
\toprule
&       & \multicolumn{9}{c}{\# photons sent $m$}\\
&  $P_{nm}$     & 0 & 1 & 2 & 3 & 4 & 5 & 6 & 7 & 8 \\
\midrule
\multirow{9}[0]{*}{\begin{sideways} \# photons detected $n$ \end{sideways}} & 0 & 1.0000	&	0.1050	&	0.0110	&	0.0012	&	0.0001  &	0.0000	&	0.0000	&	0.0000	& 0.0000\\
& 1 & 0	&	0.8950	&	0.2452	&	0.0513	&	0.0097	&	0.0017	&	0.0003	&	0.0001	& 0.0000\\
& 2 & 0	&	0	&	0.7438	&	0.3770	&	0.1304	&	0.0384	&	0.0104	&	0.0027	&	0.0007\\
& 3 & 0	&	0	&	0	&	0.5706	&	0.4585	&	0.2361	&	0.0996	&	0.0375	&	0.0132\\
& 4 & 0	&	0	&	0	&	0	&	0.4013	&	0.4672	&	0.3346	&	0.1907	&	0.0952	\\
& 5 & 0	&	0	&	0	&	0	&	0	&	0.2565	&	0.4076	&	0.3870	&	0.2862\\
& 6 & 0	&	0	&	0	&	0	&	0	&	0	&	0.1476	&	0.3066	&	0.3724\\
& 7 & 0	&	0	&	0	&	0	&	0	&	0	&	0	&	0.0755	&	0.1985	\\
& 8 & 0	&	0	&	0	&	0	&	0	&	0	&	0	&	0	&	0.0338 \\
\bottomrule
\end{tabular}
\end{center}
\end{table}

\section{Conclusion}
\label{sect:conclusion}

In conclusion, we demonstrate an SNSPD array composed of 14 individual pixels with a maximum system detection efficiency (SDE) of 90\% and an average single-pixel jitter of $\sim$\SI{20}{\pico\second}. The detector is able to operate with SDE above 80\% up to \SI{400}{\mega\cps}, and shows a maximum count rate (MCR) of \SI{1.5}{\giga\cps}. These capabilities open up new prospects for single-photon applications, such as QKD, that require the highest efficiency detection at elevated detection rates \cite{grunenfelder2023fast}. In fact, a similar device, that we fabricated and characterized, has already been exploited to demonstrate high-speed QKD, with secret-key rates exceeding \SI{60}{\mega\bps} over a distance of \SI{10}{\kilo \meter}~\cite{grunenfelder2023fast}. We investigate in depth the dynamic PNR capabilities of the array and demonstrate accurate state reconstruction for different photon-number statistics for a wide range of light inputs, imitating operation with photons with long coherence time, as is of interest for LOQC, GBS, and quantum repeater protocols. We show a 2\nobreakdash-photon fidelity of $74\%$ and $57\%$ for a 3\nobreakdash-photon state, which represent state-of-the-art results for fiber-coupled SNSPDs. Such detectors could find immediate application in LOQC protocols where the capability to distinguish few photon-number states is sufficient -- that is, either `1' \textit{vs} `more than 1 photons'~\cite{o2007optical, ralph2010optical}. Scaling up the number of pixels and boosting the SDE even further, would allow these detectors to potentially compete with transition edge sensors (TES) in GBS experiments involving many-photon states at telecom wavelength~\cite{arrazola2021quantum, madsen2022quantum}, while allowing operations at much higher rates than what is currently possible.

\newpage

\begin{suppinfo}

Supporting Information. Detailed fabrication process, description of measurements apparatus, PNR model, scaling of PNR performances. 

\end{suppinfo}

\begin{acknowledgement}  We acknowledge financial support from the Swiss NCCR QSIT, from the SNSF Practice-to-Science Grant No.~199084 and Grant No.~200020-182664, from Innosuisse Grant No.~40740.1 IP-ENG and from NRC CSTIP grant No.~QSP043. L.S.~is part of the AppQInfo MSCA ITN which received funding from the EU Horizon 2020 research and innovation program under the Marie Sklodowska-Curie grant agreement No.~956071. We acknowledge the support of Claudio Barreiro for the design and test of several electronic components of the biasing and read-out circuit. We acknowledge the assistance of the technical staff at CMi, EPFL, and of Riad Berrazouane during the fabrication of the devices. We also acknowledge the help of Angelo Gelmini in the set-up of the laser and the electro-optical modulator for the PNR experiments. We thank Adrien Python for his assistance on the use of the data acquisition software. 
\end{acknowledgement}

\begin{itemize}
\item Conflict of interest/Competing interests: The authors declare no conflicting interests. 
\item Availability of data and materials:
The data that support the findings of this study are available from the corresponding author upon reasonable request.
\item Authors' contributions:
M.P., G.V.R. H.Z. and F.B. conceptualized the project. G.V.R, L.S. and M.P. carried out the characterization of the detectors and all the experiments. G.V.R., L.S and T.B performed the the data analysis and generated the graphs. S.E.K. and G.V.R. fabricated the detector. H.Z., R.T. and F.B. supervised the project. G.V.R. and L.S. wrote the manuscript with contributions from all co-authors. 
\end{itemize}

\noindent

\bibliography{references}
\end{document}


\section{Detailed fabrication process}\label{sec:fab}
The fabrication process starts with \SI{100}{\mm} silicon wafers with \SI{100}{\nm} dry silicon oxide, which is grown by the CMi at EPFL. We evaporate metallic mirrors using a lift-off process with LOR/AZ1512 photoresist. We use direct laser writing to expose the wafers. We then sputter the SiO$_2$ needed to form the optical cavity from a SiO$_2$ target and in an Ar/O$_2$ atmosphere. The optical cavity thickness is optimized via COMSOL simulations, in order to maximize absorption of photons at \SI{1550}{\nm} in the superconducting layer. The NbTiN superconducting layer is deposited using magnetron sputtering from a single NbTi target in a nitrogen-rich atmosphere, with the quality of the layer is then verified with AFM roughness measurements and resistivity measurements. The nanowires are patterned using electron-beam lithography with a 100~keV tool and CSAR ebeam resist. The nanowires are etched with reactive ion-etching (RIE), using SF$_6$ gas. Subsequently Ti/Au electrodes are evaporated with the same lift-off process previously described. The final step of the fabrication process is a standard the deep-reactive ion etching (DRIE) Bosch process that allows the release of the detectors from the wafer. The detectors are then packaged on custom-made holders and wire-bonded to a custom-designed PCB.

\section{Measurement setup}\label{sec:meas}

\begin{figure}[ht!]
	\includegraphics[width = 0.8
	\columnwidth]{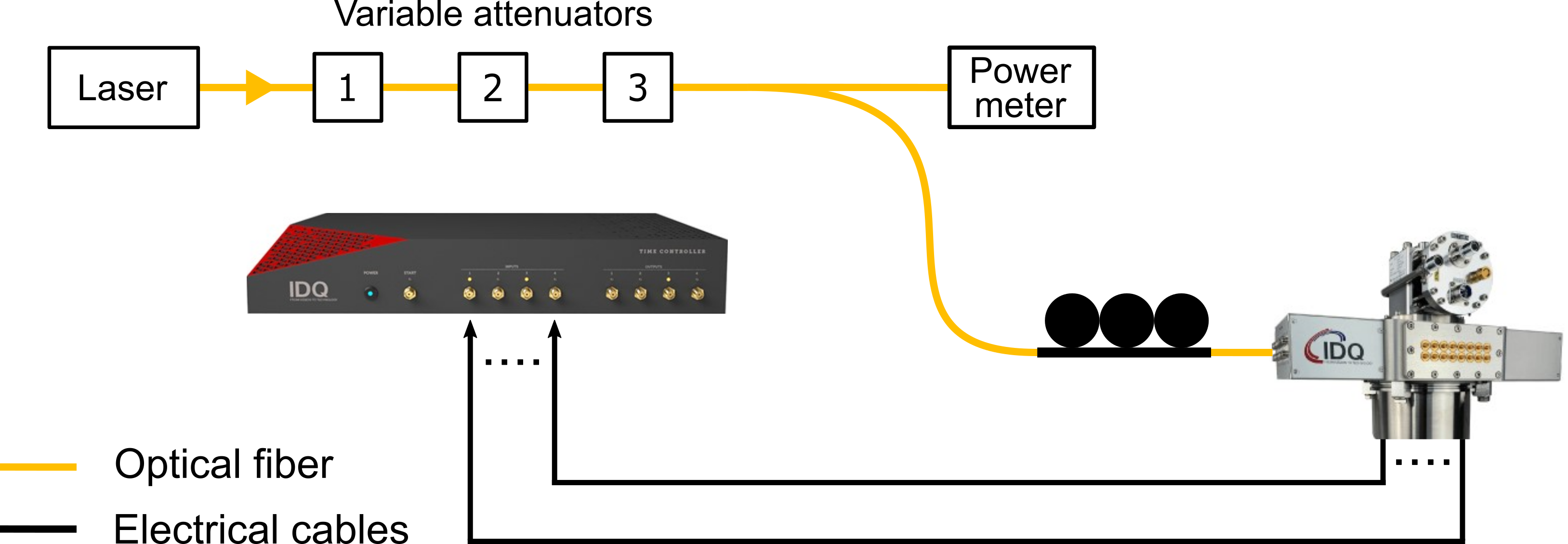}
	\caption{\label{fig:1pix} \textbf{Schematic description of the setup used for the optical characterization. Logo credit: ID Quantique.}}
\end{figure}

The detectors are placed in a 16 channel closed-cycle cryostat (ID281, ID Quantique), that can reach temperatures down to \SI{0.8}{\K}. Each pixel is biased independently using a custom-made board with 14 isolated voltage sources. The RF pulse generated by the SNSPD upon a photon detection is amplified at \SI{40}{\K} with an AC-coupled custom-made amplifier board and then amplified again at \SI{300}{\K} with a commercially available amplifier (Minicircuit ZFL-1000LN). For the efficiency, count rate, and recovery time measurements of each individual pixel we use a continuous wave (CW) laser at \SI{1550}{\nm}(EXFO FLS-2600B), attenuated with 3 optical attenuators (EXFO FVA-3150) to the single-photon level (around 1 million photons per second). Each attenuator is calibrated at each measurement using a powermeter (EXFO PM-1100). A manual polarization controller is used in order to control the polarization of the photons. The jitter of each pixel is measured at \SI{1550}{\nm} using a picosecond fiber-laser (Nuphoton technologies) in combination with a time-correlated single photon counting (TCSPC) card (B\&H). All the thermal-crosstalk measurements are also acquired using the picosecond laser. The discrimination and time-tagging of the counts is performed with time controllers (ID1000, ID Quantique). When measuring the count rate of the entire array (14 pixels) we use 4 ID1000 time-controllers. When looking at the PNR capabilities of our SNSPD array, the 4 synchronized ID1000 time-controllers are connected together and synchronized with each other. This allows us to simultaneously acquire the time-stamps for all 14 channels simultaneously. By post-processing the acquired timestamps, we obtain the number of counts for each $n$\nobreakdash-click event. By normalizing them with respect to the total number of events of the measurement (which are the product betweeen the repetition rate and the acquisition time) we can obtain the experimental photon counting statistics $Q_n$.  

\section{PNR model}\label{sec:pnr_model}
The \textbf{P} matrix is estimated by using the following equation~\cite{fitch2003photon}:
\begin{equation}
    P_{nm} = \binom{N}{n} \sum_{j=0}^{n} (-1)^j \binom{n}{j}\left[(1-\eta) + \frac{n-j}{N}\eta\right]^m.
\end{equation}
It consist of combinatorial calculations based on an array of N pixels with a total system detection efficiency $\eta$; $n$ represents the number of detected photons while $m$ the number of photons impinging on the array. In order to use this model for our device we assume that the light is uniformly distributed among the pixels, \textit{i.e.} each pixel has almost the same efficiency. In our case there is only a slight efficiency variation between the pixels, so this assumption holds and the model correctly describes the operation of the array.
\newpage
\section{Scaling of PNR performances}\label{sec:scaling}
The \textbf{P} matrix description of the PNR properties of the device is used to study how photon fidelities scale with respect to number of pixels and total system detection efficiency (SDE)\cite{dauler2009photon, vetlugin2023photon}.

The photon fidelities increase both with number of pixels and total SDE, however for a fixed SDE the fidelities tend to saturate for large number of pixels \cite{vetlugin2023photon}. Considering the 2-photon fidelity $P_{22}$ we have that with a few pixels (\textit{i.e.} less than 10) is convenient to increase their numbers rather than the SDE, whereas for many pixels (\textit{i.e.} more than 10) is usually better to improve the SDE rather than the number of pixels (see Tab. \ref{table:p22}). Similar considerations apply for higher photon-number fidelities, but with larger number of pixels. 

\begin{table}
\small
\begin{center}
\caption{Variation of {$P_{22}$} with respect to overall SDE and number of pixels}\label{table:p22}%
\begin{tabular}{c c | c c c c c}
\toprule
&       & \multicolumn{5}{c}{Number of pixels}\\
&  $P_{22}$(\%)     & 4 & 8 & 14 & 32 & 50\\
\midrule
\multirow{2}[0]{*}{SDE (\%)} & 90 & 60 & 70 & 74 & 78.5 & 79.5\\
& 95 & 67 & 79 & 84 & 87.5 & 88.5 \\
\bottomrule
\end{tabular}
\end{center}
\end{table}
For reference, in Tab. \ref{table:povm_32_pixel} we report the P matrix (truncated at 10-photon events) of a possible 32-pixel array SNSPD with a total SDE of 95\%.
\begin{table}
\footnotesize
\begin{center}
\caption{Truncated $\mathbf{P}$ matrix (up to 10-photon events) of a 32-pixel interleaved array.}\label{table:povm_32_pixel}%
\begin{tabular}{c c | c c c c c c c c c c c}
\toprule
&       & \multicolumn{11}{c}{\# photons sent $m$}\\
&  $P_{nm}$     & 0 & 1 & 2 & 3 & 4 & 5 & 6 & 7 & 8 & 9 & 10\\
\midrule
\multirow{11}[0]{*}{\begin{sideways} \# photons detected $n$ \end{sideways}} & 0 & 1	&	0.0500	&	0.0025	&	0.0001	&	0.0000  &	0.0000	&	0.0000	&	0.0000	& 0.0000 & 0.0000 & 0.0000 \\
& 1 & 0	&	0.9500	&	0.1232	&	0.0122	&	0.0011	&	0.0001	&	0.0000	&	0.0000	& 0.0000 & 0.0000 & 0.0000 \\
& 2 & 0	&	0	&	0.8743	&	0.2090	&	0.0341	&	0.0047	&	0.0006	&	0.0001	&	0.0000 & 0.0000 & 0.0000 \\
& 3 & 0	&	0	&	0	&	0.7787	&	0.2944	&	0.0713	&	0.0141	&	0.0025	&	0.0004 & 0.0000 & 0.0000 \\
& 4 & 0	&	0	&	0	&	0	&	0.6704	&	0.3666	&	0.01233	&	0.0330	&	0.0077	& 0.0016 & 0.0003 \\
& 5 & 0	&	0	&	0	&	0	&	0	&	0.5573	&	0.4153	&	0.1849	&	0.0641 & 0.0191 & 0.0052 \\
& 6 & 0	&	0	&	0	&	0	&	0	&	0	&	0.4467	&	0.4348	&	0.2474 & 0.1078 & 0.0399 \\
& 7 & 0	&	0	&	0	&	0	&	0	&	0	&	0	&	0.3448	&	0.4245	& 0.3004 & 0.1606\\
& 8 & 0	&	0	&	0	&	0	&	0	&	0	&	0	&	0	&	0.2559 & 0.3886 & 0.3348\\
& 9 & 0	&	0	&	0	&	0	&	0	&	0	&	0	&	0	&	0 & 0.1823 & 0.3347\\
& 10 & 0&	0	&	0	&	0	&	0	&	0	&	0	&	0	&	0 & 0 & 0.1245\\
\bottomrule
\end{tabular}
\end{center}
\end{table}
\bibliography{references}